\documentclass[reprint,groupedaddress,prb]{revtex4-1}
\usepackage{amsmath,amssymb,bm}
\usepackage{braket}
\usepackage{cases}
\usepackage[usenames]{color}
\usepackage{graphicx}% Include figure files
\usepackage{dcolumn}% Align table columns on decimal point
\usepackage{ulem}
\providecommand{\noopsort}[1]{}

\begin{document}

%\preprint{APS/123-QED}

\title{Weyl points of mechanical diamond}

\author{Yuta Takahashi$^1$}
\email{takahashi@rhodia.ph.tsukuba.ac.jp}
\author{Toshikaze Kariyado$^2$}%
\email{kariyado.toshikaze@nims.go.jp}
\author{Yasuhiro Hatsugai$^3$}
\email{hatsugai.yasuhiro.ge@u.tsukuba.ac.jp}
\affiliation{%
$^1$Graduate School of Pure and Applied Science, University of Tsukuba, Tsukuba, Ibaraki 305-8571, Japan\\
$^2$International Center for Materials Nanoarchitectonics(WPI-MANA), National Institute for Materials Science, Tsukuba, Ibaraki 305-0044, Japan\\
$^3$Division of Physics, University of Tsukuba, Tsukuba, Ibaraki 305-8571, Japan
}%


%\affiliation{
%\affiliation{
% Third institution, the second for Charlie Author%}%
%\affiliation{%

\date{\today}% It is always \today, today,

\begin{abstract}
%\textcolor{cyan}{A spring-mass model arranged in a diamond structure ---mechanical diamond---has
%attractive topological properties induced by a symmetry breaking.
%We introduce additional springs connecting the next-nearest-neighboring pairs of
%mass points and the tuning of the mass parameter to the existing mechanical diamond,
%which generates several Weyl points in the frequency dispersion.
%A mapping of the Weyl points modulated by a uniform outward tension
%is discussed from the tetrahedral symmetry of the NNN springs.
%There happens a transmutation of the monopole charges rapidly
%as the tension changes. Fermi arcs in an anisotropic system
% with a boundary is also performed.}
 A spring-mass model arranged in a diamond
 structure --- mechanical diamond --- is analyzed in terms of
 topology in detail. We find that, additional springs connecting the
 next-nearest-neighbor pairs of mass points and the modulation of the
 mass parameters to the pristine mechanical diamond generates multiple
 pairs of Weyl points in the frequency dispersion. Evolution of the Weyl
 point positions in the Brillouin zone against the uniform outward tension
 is tracked and explained by the point group symmetry,
 especially tetrahedral symmetry of the NNN springs. Interestingly,
 there happens a rapid transmutation of the monopole charges of the Weyl
 points as the tension varies. We also show surface Fermi arcs
 in the case with anisotropy in the NNN springs.
 
\end{abstract}

\maketitle

%\tableofcontents

\section{\label{sec:intro}Introduction}
%\textcolor{cyan}{
%Topological semimetal has been discussed recently in both a theoretical and a experimental points of view,
%including Dirac semimetal and nodal line semimetal~\cite{Young2012,Wang2012,Liu2014,Burkovnodal2011,kobayashi2017,Yan2017,Bian2016}.
%These semimetals have topological gapless points or lines protected by a symmetry, in short.
%On the contrary, Some semimetals are generated by a symmetry breaking.
%Weyl point is a representative example with isolated gapless points~\cite{Murakami2007,BurkovWeyl2011,Fang2012,Soluyanov2015,Okugawa2017,Lu2015}.
%The gapless points are created/annihilated in even number of pairs and have a $\pm 1$ topological charge respectively, but
%only under certain conditions that the sum of the charges is preserved.
%Topological stability actually stems from this monopole charge.
%A Weyl point is never vanished unless the point encounters another with the opposite chirality.
%Fermi arc, a intersection line of chiral edge mode and the Fermi energy that connects a pair of Weyl points,
%is also a noteworthy feature~\cite{Wan2011,XuFer2015}. Note that the edge mode is characterized by the bulk Chern number, therefore
%this is one instance of bulk-edge correspondence~\cite{Hatsugai1993}.}

%\textcolor{red}{
Topological semimetals, which are characterized by isolated gapless
points or lines in their band structure, have been intensively studied
both in theoretical and experimental view points recently. There are two
classes of topological semimetals: one requires some symmetry for
protection of the gapless points, while the other is purely
topological. The former includes Dirac semimetals with time reversal and
spatial inversion symmetries, or nodal line semimetals that are protected
by reflection symmetry, for instance \cite{Young2012,Wang2012,Liu2014,Burkovnodal2011,kobayashi2017,Yan2017,Bian2016}. 
On the other hand, the latter includes Weyl semimetals where either of
time reversal or spatial inversion symmetry is requried to be broken
\cite{Murakami2007,BurkovWeyl2011,Fang2012,Soluyanov2015,Okugawa2017,Lu2015}.
Namely, for the Weyl semimetals, symmetry breaking is the key for generating
gapless points (Weyl points). The Weyl points are created/annihilated as
pairs and each of a pair carries $\pm 1$ topological charge, and the
conservation of the topological (or monopole) charge is behind the
stability of the Weyl points. The topological nature of the Weyl points are also reflected in surface states: a Weyl semimetal is characterized by a Fermi arc, which is
an isoenergy line of the surface state at the Fermi energy connecting 
a pair of Weyl points projected onto the surface Brillouin zone
\cite{Wan2011,XuFer2015}. We can relate the surface Fermi arc with the bulk Chern number, forming a typical example of bulk-edge correspondence~\cite{Hatsugai1993}.
%}

%\par
%\textcolor{cyan}{Another hot topic in condensed matter physics is classical system discussed from a topological perspective
%such as photonics~[\onlinecite{Raghu2008,Wang2008}] and mechanics~[\onlinecite{Prodan2009,Kane2014,Po2016,Paulose2015,Rocklin2016,Wang2015,Joshua2017}].
%Here we can see topological edge modes as in a quantum system,
%though there are some differences. The most significant distinction is its controllability.
%In particular, its nature is remarkable in the topological mechanics.
%The artifical structure enables us to implement any modification.
%Mechanical graphene, a spring-mass model with the honeycomb structure, is the simplest model among them~[\onlinecite{Cserti2004,WangYao2015,Kariyado2015}].
%A frequency dispersion of this system has Dirac cones that change those numbers and positions by a uniform outward tension.
%Chiral edge states are also induced by a time-reversal symmetry breaking.
%Furthermore, We have proposed a diamond spring-mass model---mechanical diamond---as an analogue in 3D~[\onlinecite{takahashi2017}].
%A chiral symmetry give rise to line nodes and multiple zero mode edge
%states.}

%\textcolor{red}{
Currently, the concept of topological insulators/semimetals and bulk-edge correspondence is extended to wider and wider variety of
systems including classical systems such as photonic crystals
\cite{Raghu2008,Wang2008} and mechanical systems
\cite{Prodan2009,Kane2014,Po2016,Paulose2015,Rocklin2016,Wang2015,Joshua2017}. It
is possible to mimic quantum Hall states, quantum spin Hall states, and
other topologically nontrivial states by appropriate artificial
designs. Roughly speaking, the typical design goes as follows. In a spatially periodic classical system, the frequency-wavenumber dispersion relation of the normal modes forms a band structure, just like the energy-momentum dispersion relations in quantum solids. Then, we choose structures and parameters so as to make the band structure of the normal modes resembling to the band structure of topological insulators/semimetals. The designed nature is an advantage of the classical topological systems over the quantum electronic systems. Namely, the classical systems usually have better controllability than the quantum
ones, and enables us to access parameter regions that are impossible or
difficult in experiments in quantum systems. 
%\textcolor{red}{This mechanical property becomes a foothold to
%investigate that of solid states}
Mechanical graphene, a spring-mass model
with the honeycomb structure having Dirac cones in its frequency
dispersion, enjoys this controllable feature
\cite{Cserti2004,WangYao2015,Kariyado2015}. For instance, the number and
positions of the Dirac cones can be changed simply by a uniform outward
tension. If the Dirac cones are gapped out by a time-reversal
symmetry breaking term, there appear chiral edge mode, and the uniform outward tension causes a flip in the flowing direction of the chiral edge
mode. Mechanical graphene has a natural extension in three-dimension,
i.e., mechanical diamond that is a spring-mass model with diamond structure
\cite{takahashi2017}. We have found that mechanical diamond is a
classical counterpart of the nodal line semimetal, i.e., having lines of degeneracy in the frequency dispersion, and a (modified)
chiral symmetry is the key for line node protection\cite{takahashi2017}.
%}

%\par
%\textcolor{cyan}{In the present work, we extend the concept of the mechanical diamond by connecting next-nearest-neighbor
%springs supplementally and changing a mass of mass points to break the chiral symmetry.
%The symmetry breaking vanishes the line nodes and generates some Weyl points moving along the former line nodes.
%We discuss that the number of the Weyl points is determined by a tetrahedral symmetry of the added springs. 
%A chirality change happens during the movememt of the Weyl points.
%We also refer to Fermi arc induced by a nontrivial topology in an anisotoropic system.
%This paper is arranged as follows. In Sec.\ref{sec:model}, we introduce two types of the mechanical diamond,
%a nearest-neighbor model and a next-nearest-neighbor model. In Sec.\ref{sec:method}, we explain
%how to examine Weyl points numerically. Mappings of the Weyl points for a isotropic case
%and an anisotropic case are shown in Sec.\ref{sec:pos_NNN}. The result is discussed
%by a symmetry. Sec.\ref{sec:Fer} considers Fermi arc for the anisotropic case with a boundary.
%We give a conclusion in Sec.\ref{sec:con}.
%}

%\textcolor{red}{
In this paper, we investigate symmetry breaking effects
on the frequency-wavenumber dispersion relation of the normal vibration modes in the mechanical diamond. The (modified) chiral symmetry is broken by
introducing springs connecting next-nearest-neighbor (NNN) pairs of mass
points in addition to the original ones for the nearest neighbor (NN)
pairs, and introducing modulation in mass of the mass points. 
This symmetry breaking eliminates the line nodes and generates several
Weyl points.  It is
noticed that the tetrahedral symmetry of the system is crucial in determining the number of the Weyl points. As
in the case of mechanical graphene, the uniform outward
tension induces movement of the Weyl points along the line originally
being line nodes. Interestingly, an exchange of chirality (or
topological charge) occurs as the Weyl points travel about. We also
investigate surface states to see Fermi arcs associated with the
nontrivial topology in the bulk with Weyl points. 

In the following, we
start with describing two types of the mechanical diamond,
a NN model and a NNN model in
Sec.~\ref{sec:model}. Section~\ref{sec:method} is for numerical methods
to identify Weyl points. Mappings of the Weyl points for an isotropic case
and an anisotropic case are shown in Sec.~\ref{sec:pos_NNN}, and the
results are interpreted in terms of symmetry. Section~\ref{sec:Fer}
considers Fermi arcs for the anisotropic case in a system with a
boundary. The paper is concluded in Sec.~\ref{sec:con}.
%}

\section{\label{sec:model}model}

\begin{figure*}[tbp]
 \centering
  \includegraphics[width=16cm]{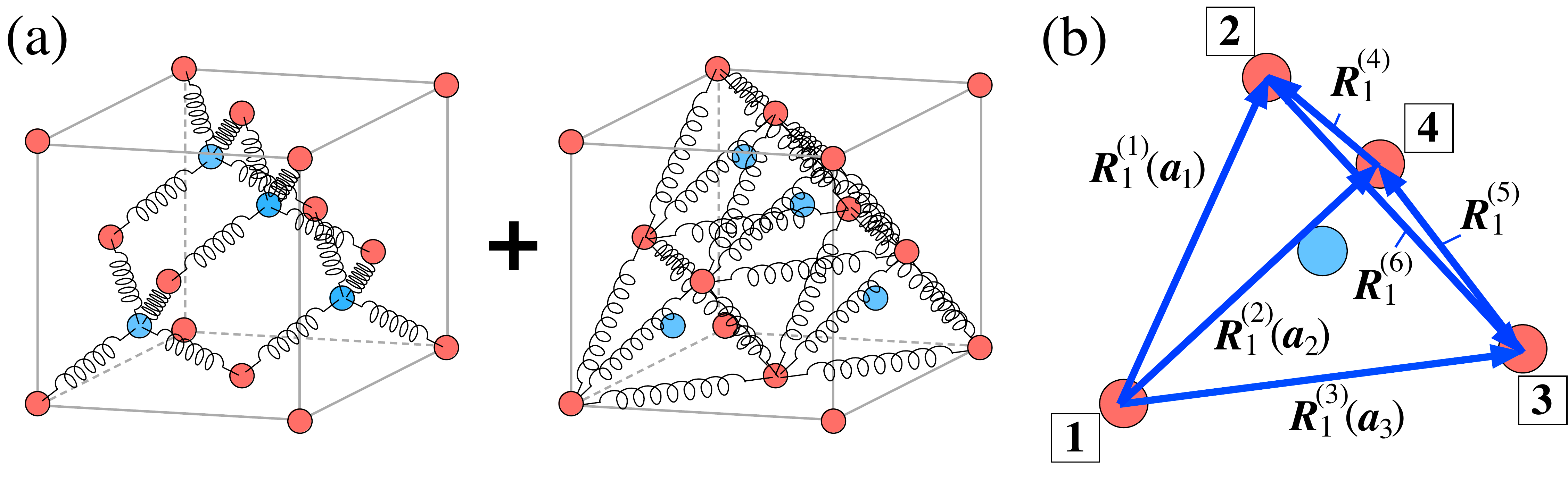}
\caption{\label{fig:NNN}{(a) Schematic picture of the NNN model. Red (blue) mass points are labeled as sublattice A (B).
(b) Definitions of NNN vectors $\bm R_1^{(i)}$. In particular, the three vectors $\bm R_1^{(1)}$-$\bm R_1^{(3)}$
are exactly the same as the unit translation vectors $\bm a_1$-$\bm a_3$
 respectively. The four vertices of a tetrahedron formed by the NNN
 springs are numbered (from 1 to 4) to discuss a symmetry in \S\ref{sec:pos_NNN}.}}
\end{figure*}

\subsection{Nearest-neighbor (NN) model}
%A mechanical diamond is a simple classical model composed of mass points
%and springs in a diamond structure. The springs connect the
%neighboring pairs of the mass points. We use some parameters such as
%spring constant of the NN springs $\kappa$, mass of the mass points $m$, natural length
%of the NN springs $l_0$ and distance between the neighboring mass points $R_0$. 
%By a uniform outward tension, the distance $R_0$ becomes longer compared to the natural length $l_0$. 
%This deformation is allowed in a proper boundary condition.
%The mass $m$ is set to a unit in this NN model.
%Given that the oscillation of the mass points is close to the equilibrium point,
%we adopt a variable of the displacement of each mass point as $\bm x_{\bm R a}=(x_{\bm R a}, y_{\bm R a}, z_{\bm R a})$.
%Here, $\bm R$ denotes the lattice points and $a$ is a sublattice index.

A mechanical diamond is a classical
model composed of mass points
and springs arranged in a diamond
structure. We first consider NN model, where the
springs connect the
neighboring pairs of the mass points.
The parameters describing our model are
spring constant of the NN springs $\kappa$, mass of the mass points $m$, natural length
of the NN springs $l_0$ and distance between the neighboring mass points $R_0$. 
If we apply uniform
outward tension, the distance $R_0$ 
can be larger than the natural length $l_0$. 
This deformation is 
enabled by a proper boundary condition. 
One possible way to exert the outward tension is to hang the entire system into some cage, 
and to apply the fixed boundary condition by connecting the outer most mass points to the cage.
For simplicity, $m$
is set to unity in this
NN model. 
The dynamical variables of the model are
$\bm{x}_{\bm{R}a}=(x_{\bm{R}a}, y_{\bm{R}a}, z_{\bm{R}a})$, which are
displacements of mass points from the equilibrium positions
with $\bm{R}$ denoting the lattice points and $a$ being a sublattice
index. In the following, we focus on infinitesimal oscillation about
the equilibrium positions, and $\bm{x}_{Ra}$ is regarded as a small
quantity.

%The elastic energy of a linear spring $U_s$ can be expressed as
%$U_s \simeq U_0+U_1(\delta x_{\lambda})+U_2(\delta x_{\mu} \delta x_{\nu})$,
%with $\delta \bm x$ indicating a difference between the displacements of
%the neighboring pair of the mass points. As a matter of fact, we simply focus on $U_2(\delta x_{\mu} \delta x_{\nu})$ 
%to solve the Newton's equation around the equilibrium point. This key term $U_2(\delta x_{\mu} \delta x_{\nu})$ is written as
%\begin{equation}
% U_2(\delta x_{\mu} \delta x_{\nu})=\frac{1}{2}\kappa\, \delta x_{\mu} \gamma_{\bm R_0}^{\mu \nu} \delta x_{\nu},
%\end{equation}
%where $\gamma^{\mu \nu}_{\bm R_0}=(1-\eta)\delta^{\mu \nu}+\eta \hat{R}_0^{\mu} \hat{R}_0^{\nu}$ and
%$\hat{\bm R}_0=\bm R_0/|\bm R_0|$, $\eta \equiv l_0/R_0$. We take the summation over $\mu$ and $\nu$,
%which deplete the three directions of the displacement $x$, $y$ and $z$.
%From the term, we can find out that the oscillation is characterized by
%the tension parameter $\eta$.

The elastic energy of a single linear spring $U_s$
is approximated as 
$U_s \simeq U_0+U_1(\delta x_{\lambda})+U_2(\delta x_{\mu} \delta x_{\nu})$,
with $\delta \bm x$ indicating a difference between the displacements of
the neighboring pair of the mass points. The $U_1$-term
that is first order in $\delta x_\lambda$ can be finite for a single
spring, but if we consider all the springs connected to a single mass
point, the terms cancel out with each other. Then, the dynamics of the
system is governed by the $U_2$-term that is second order in $\delta
x_\lambda$. The $U_2$-term is explicitly written as
\begin{equation}
 U_2(\delta x_{\mu} \delta x_{\nu})=\frac{1}{2}\kappa\, \delta x_{\mu} \gamma_{\bm R_0}^{\mu \nu} \delta x_{\nu},
\end{equation}
where $\gamma^{\mu \nu}_{\bm R_0}=(1-\eta)\delta^{\mu \nu}+\eta \hat{R}_0^{\mu} \hat{R}_0^{\nu}$,
$\hat{\bm R}_0=\bm R_0/|\bm R_0|$ and $\eta \equiv l_0/R_0$.
The summation over $\mu$ and $\nu$, which run
through three spatial directions ($x$, $y$, $z$), is implicitly taken.
Importantly, the tension parameter $\eta$ determines the
$U_2$-term.

Substituting four NN vectors
$\bm{R}_0^{(i)}$ ($i=1,2,3,4$) into $U_2(\delta x_{\mu} \delta
x_{\nu})$, and applying Fourier transformation in time and space, i.e.,
using $\bm x_{\bm R a} =\frac{1}{N}\sum_{\bm
k}\mathrm{e}^{\mathrm{i}(\bm k\cdot \bm R+\omega t)}\bm \phi_{\bm k a}$,
the Newton's equation of motion becomes
\begin{equation}
\hat{\Gamma}(\bm k)\bm{\phi}(\bm k)=\omega^2\bm{\phi}(\bm k),
\end{equation}
where 
\begin{equation}
 \hat{\Gamma}(\bm k)=4\kappa(1-\frac{2}{3}\eta)\hat{1}+
\begin{pmatrix}
\hat{0} & \hat{\Gamma}_{AB}(\bm k)\\
\hat{\Gamma}^{\dagger}_{AB}(\bm k) & \hat{0}
\end{pmatrix},
\end{equation}
and $\hat{\Gamma}_{AB}(\bm k)=-\kappa (\hat{\gamma}_4+\mathrm{e}^{\mathrm{-i}\bm k \cdot \bm a_1}\hat{\gamma}_1
+\mathrm{e}^{\mathrm{-i}\bm k \cdot \bm a_2}\hat{\gamma}_2+\mathrm{e}^{\mathrm{-i}\bm k \cdot \bm a_3}\hat{\gamma}_3)$,
with $\hat{\gamma}_i\equiv\hat{\gamma}_{\bm R_0^{(i)}} \ (i=1,2,3,4)$. 
A frequency dispersion
is obtained from this secular equation.
Note that $\hat{\Gamma}(\bm k)$ has a chiral symmetry
if the diagonal terms
are subtracted, 
i.e., $\hat{\Gamma}'(\bm k)=\hat{\Gamma}(\bm
k)-4\kappa(1-\frac{2}{3}\eta)$ anticommutes with
$\hat{\Upsilon}=\mathrm{diag}(1,1,1,-1,-1,-1)$. 
As discussed in
Ref.~\onlinecite{takahashi2017}, by making use of the Berry's
parameterization of a two band effective model near a gap closing point,
it can be shown that the gap closing point should form a line (line
node) in the 3D Brillouin zone \cite{Blout1985} with the chiral
symmetry. 
In contrast, the gap closing point
becomes an isolated point (Weyl point) in the 3D Brillouin zone without chiral
symmetry.
In the following, we consider
to add a term breaking the chiral symmetry in mechanical diamond.

\subsection{Next-nearest-neighbor (NNN) model}

In order to break the chiral symmetry, we introduce (i)
springs connecting the next-nearest-neighbor pairs of the mass points on
one of the sublattices only (say, the sublattice A out of the
sublattices A and B) (see Fig.~\ref{fig:NNN}), and (ii) difference in
the mass of the mass points between the two sublattices by writing the mass for the sublattice A as $m$ while keeping the mass for the sublattice B to be 1. Following the
arguments in the previous subsection, and using six NNN
vectors $\bm{R}^{(i)}_1$ ($i=1,2,...,6$) in Fig.~\ref{fig:NNN}, the
secular equation to be solved becomes
\begin{equation}
 \hat{\Gamma}_{NNN}(\bm k)\bm \phi'(\bm k)=\omega^2\bm \phi'(\bm k),
\end{equation}
with
\begin{equation}
\hat{\Gamma}_{NNN}(\bm k)=
\begin{pmatrix}
 \hat{\Gamma}_{AA}(\bm k)/m & \hat{\Gamma}_{AB}(\bm k)/\sqrt{m}\\
 \hat{\Gamma}_{AB}^{\dagger}(\bm k)/\sqrt{m} & \hat{\Gamma}_{BB}
\end{pmatrix}.
\end{equation}
Here,
\begin{widetext}
\begin{eqnarray}
 \hat\Gamma_{AA}(\bm k)=&&\hat{\Gamma}_{BB}
+2\kappa'_1[1-\cos({\bm k \cdot \bm a_1})]\hat{\gamma}'_1
+2\kappa'_2[1-\cos({\bm k \cdot \bm a_2})]\hat{\gamma}'_2
+2\kappa'_3[1-\cos({\bm k \cdot \bm a_3})]\hat{\gamma}'_3\nonumber \\
&&+2\kappa'_4[1-\cos({\bm k \cdot(\bm a_1-\bm a_2)})]\hat{\gamma}'_4
+2\kappa'_5[1-\cos({\bm k \cdot(\bm a_2-\bm a_3)})]\hat{\gamma}'_5
+2\kappa'_6[1-\cos({\bm k \cdot(\bm a_3-\bm a_1)})]\hat{\gamma}'_6,
\end{eqnarray}
\end{widetext}
with $\hat{\gamma}'_i\equiv\hat{\gamma}_{\bm R^{(i)}_1}=(1-\eta')\delta^{\mu \nu}+\eta' \hat{R}_1^{(i)\mu} \hat{R}_1^{(i)\nu}$,
$\hat{\bm R}_1=\bm R_1/|\bm R_1|$, $\hat{\Gamma}_{BB}=4\kappa(1-\frac{2}{3}\eta)\hat{1}$ and $\hat{\Gamma}_{AB}(\bm k)$
is unchanged from the previous subsection. 
The new tension parameter $\eta'$ is defined as $\eta'\equiv l_1/R_1$ where $l_1$ is natural length
of the NNN springs and $R_1=|\bm R_1|$.
A uniform and isotropic outward tension preserves the
ratio between $R_0$ and $R_1$, and therefore, we have
$\eta'=\frac{\eta'_0}{\eta_0}\eta$ where $\eta_0$ and $\eta'_0$ are the
tension parameters of the NN and NNN springs at some reference tension,
respectively.

Note that we have assigned different spring constants
for NNN springs lying in different directions, i.e.,
$\kappa'_i$ ($i=1,2,...,6$) for
the NNN springs in the $\bm R^{(i)}_1$ directions. For
the NN springs, the forces exerted on a single mass point are balanced
between four springs in equilibrium, and it is better to assign the same
spring constants for four springs to keep the ideal diamond
structure. On the other hand, for the NNN springs, the forces on a mass
point are balanced in a pairwise way in each of the six directions along
$\bm {R}^{(i)}_1$ in equilibrium, and the ideal diamond structure is kept even
if we assign different spring constants for different directions \cite{footnote1}. As we have noted, the mass parameters depend on the sublattices, and in that case, a naive treatment makes the problem a generalized eigenvalue problem. However for convenience, we
have reformulated it into a standard eigenvalue problem with an
hermitian matrix $\hat{\Gamma}_{NNN}(\bm{k})$ by absorbing the factor
$m$ into the eigenvectors as $\bm \phi'(\bm k)=\mathrm{diag}(\sqrt m,\sqrt m,\sqrt m,1,1,1)
\bm \phi(\bm k)$. In the last subsection, subtraction of the diagonal part proportional to an identity matrix was essential to show that the model obeys a (modified) chiral symmetry. For NNN model, since the
block matrix $\hat{\Gamma}_{AA}$ depends on
the momentum, it is no longer possible to remove/neglect it in the symmetry argument.

\begin{figure}[htbp]
 \centering
  \includegraphics[width=8cm]{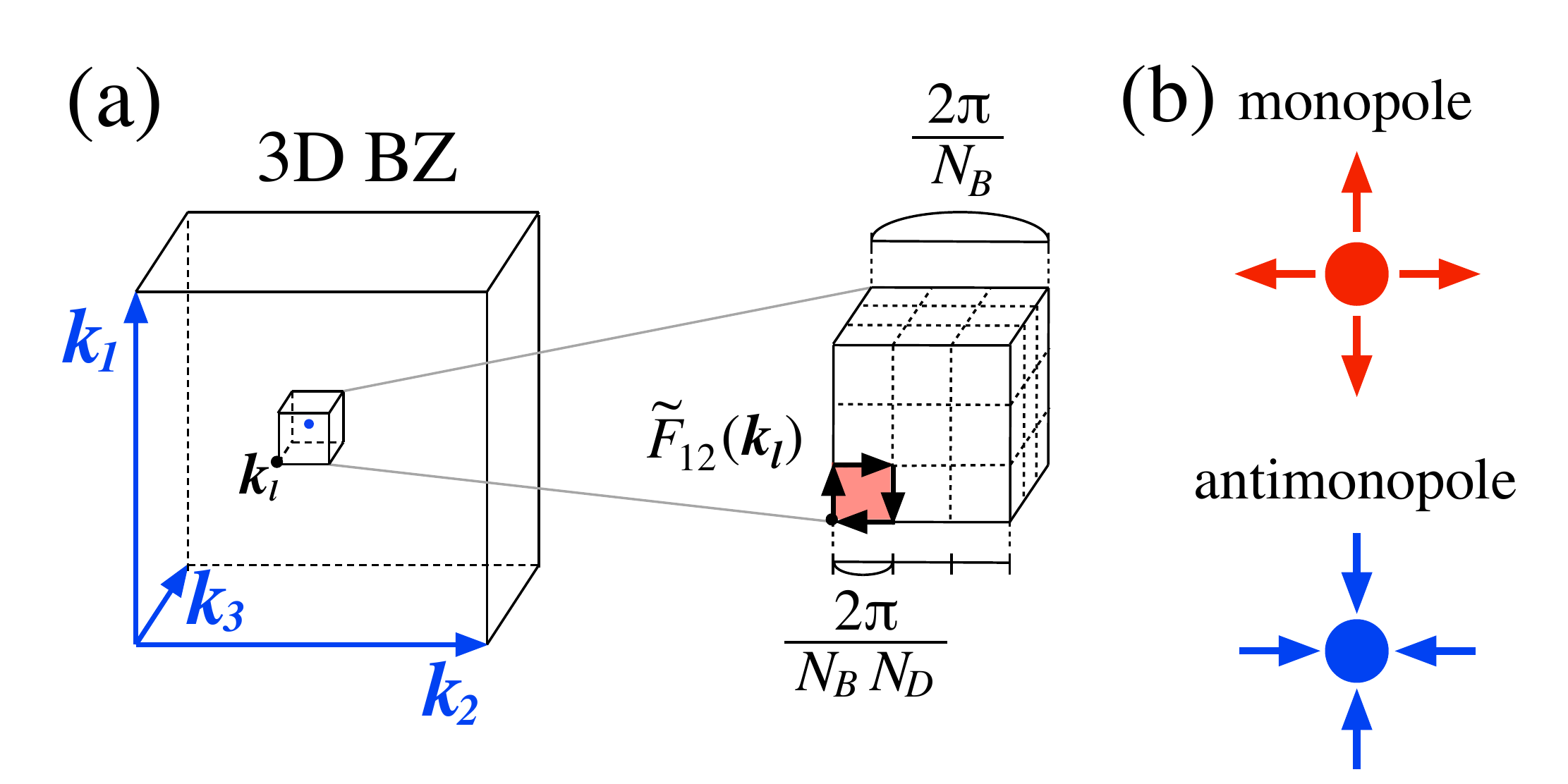}
\caption{\label{fig:WeylCalc}(a) Schematic picture of the procedure to detect Weyl points. A lattice field strength $\tilde F_{1 2}(\bm k_{\bm l})$
is calculated on the red area as a part of one face of a minicube. (b)
 Definition of a monopole (antimonopole), with the arrows represent the flow of the Berry flux. The monopole (antimonopole)
 corresponds to chirality $+$ $(-)$.}
\end{figure}

\section{\label{sec:method}numerical method}
Our major task in the following is to identify Weyl
points. In numerics, it is nontrivial to distinguish a strictly gapless
point from a tiny but finitely gapped point if we only look at
eigenvalue spectra.
Therefore, we try to extract more information from
eigenstates. Specifically, we numerically locate monopoles in Brillouin
zone (Weyl points) by evaluating Berry curvature. Here, the well-established method\cite{Fukui2005,Hirayama2017} is adapted for the evaluation of the Berry curvature. Specifically, the procedure is as follows [see also Fig.~\ref{fig:WeylCalc}(a)].
First, the Brillouin zone is
divided into coarse cubes whose corners are labeled by $\bm k_{\bm l}=(k_{l_1},k_{l_2},k_{l_3})$
where $\bm l=(l_1,l_2,l_3)$, $k_{l_\mu}=2\pi l_{\mu}/N_B$ and $l_{\mu}=0,...,N_B-1$.
Then we define a $U(1)$ link variable by
\begin{equation}
 U_{\mu}(\bm k)\equiv \frac{\mathrm{det} \,\psi^{\dagger}(\bm k) \,\psi(\bm k + \hat{\bm e}_{\mu})}
{|\mathrm{det}\,\psi^{\dagger}(\bm k)\,\psi(\bm k + \hat{\bm e}_{\mu})|},
\end{equation}
where $\psi(\bm k)$ is a triplet state of which components are three
states with lower frequencies.
We have also put in a shorthand notation $\hat{\bm{e}}_{\mu}=2\pi/(N_B N_D)(\delta_{1\mu}, \delta_{2\mu}, \delta_{3\mu})$
with a quantity $N_D$ denoting the number of discretized mesh points on the surface of the coarse cube.
For the square spanned by $\bm k_{\bm l}$, $\bm k_{\bm l}+\hat{\bm e}_{\mu}$, $\bm k_{\bm l}+\hat{\bm e}_{\nu}$
and $\bm k_{\bm l}+\hat{\bm e}_{\mu}+\hat{\bm e}_{\nu}\,(\mu\neq\nu)$, a
lattice field strength is given as
\begin{eqnarray}
 \tilde{F}_{\mu \nu}(\bm k_{\bm l}) && \equiv \mathrm{Arg} \, \frac{U_{\mu}(\bm k_{\bm l}) \, 
U_{\nu}(\bm k_{\bm l}+\hat{\bm e}_{\mu})}
{U_{\mu}(\bm k_{\bm l}+\hat{\bm e}_{\nu}) \,
U_{\nu}(\bm k_{\bm l})},\nonumber \\
&&-\pi<\tilde{F}_{\mu \nu}(\bm k_{\bm l}) \leq \pi,
\end{eqnarray}
and a monopole charge of one coarse cube $\tilde{C}_{\bm l}$ is computed from $\tilde{F}_{\mu \nu}(\bm k_{\bm l})$ as 
\begin{widetext}
\begin{eqnarray}
 \tilde{C_{\bm l}}=&&\frac{1}{2\pi}\sum_{i,j=0}^{N_D-1}\bigl[-\tilde{F}_{12}(\bm k_{\bm l}+i\hat{\bm{e}}_1+j\hat{\bm{e}}_2)
-\tilde{F}_{23}(\bm k_{\bm l}+i\hat{\bm{e}}_2+j\hat{\bm{e}}_3)
-\tilde{F}_{31}(\bm k_{\bm l}+i\hat{\bm{e}}_3+j\hat{\bm{e}}_1)\nonumber\\
+&&\tilde{F}_{12}(\bm k_{\bm l+(0, 0, 1)}+i\hat{\bm{e}}_1+j\hat{\bm{e}}_2)
+\tilde{F}_{23}(\bm k_{\bm l+(1, 0, 0)}+i\hat{\bm{e}}_2+j\hat{\bm{e}}_3)
+\tilde{F}_{31}(\bm k_{\bm l+(0, 1, 0)}+i\hat{\bm{e}}_3+j\hat{\bm{e}}_1)\bigr].
\end{eqnarray}
\end{widetext}
When this monopole charge results in a nonzero integer,
at least one Weyl point exists in that cube.
(In general, we have to note that even if this monopole
charge is zero, we still have a chance to find multiple Weyl points
with compensated charges.) Hence, we pick up cubes with
finite monopole charges and repeat this
procedure (subdivide into smaller cubes) 
until we have a solo Weyl point in each cube.
%acquire the position of the Weyl points by repeating this method with
%setting $N_B$ large enough so that the merely single Weyl point lies in the cube.  

\begin{figure}[tbp]
 \centering 
  \includegraphics[width=8.5cm]{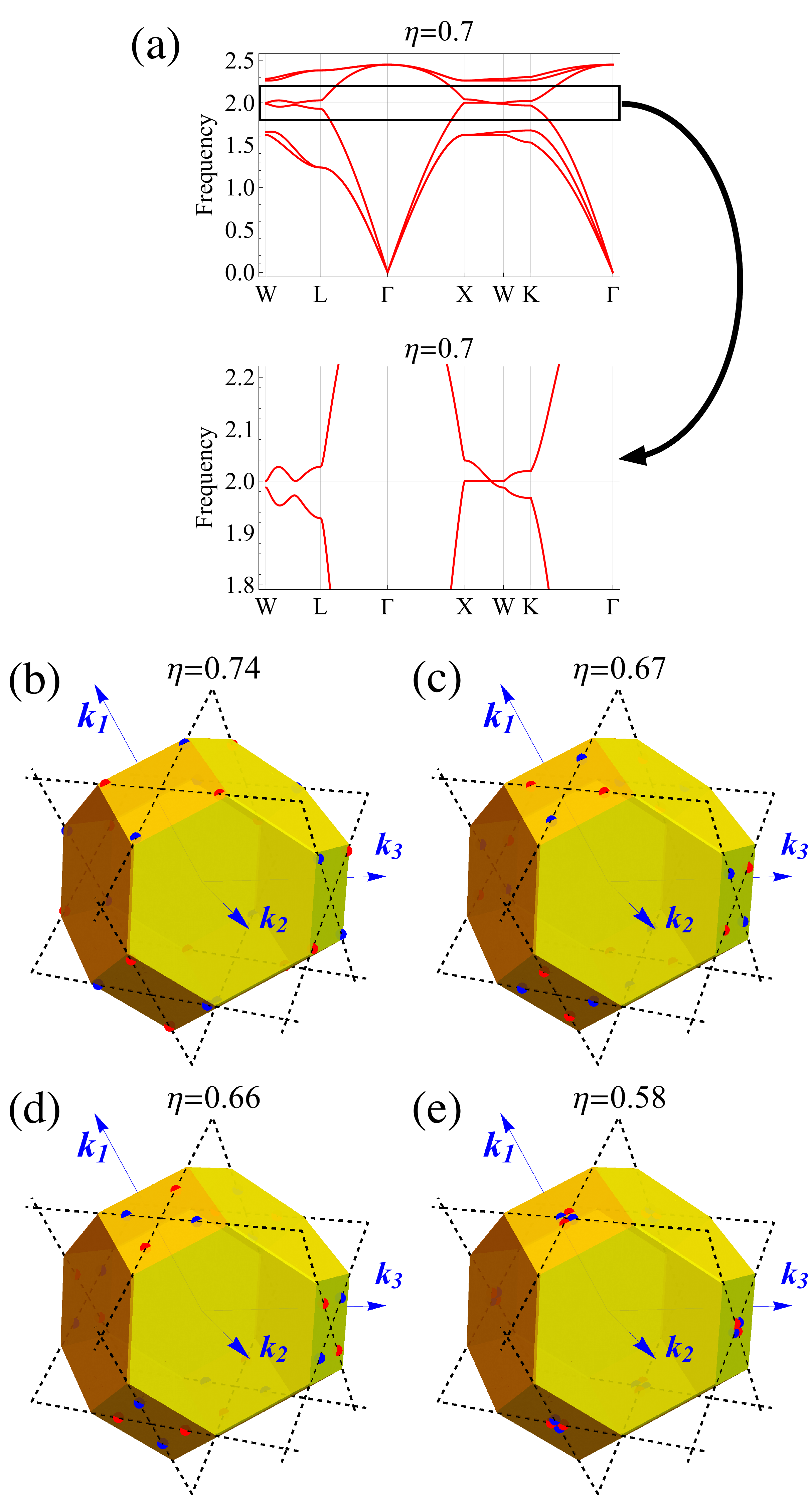}
\caption{\label{fig:WeylIso}(a) Bulk band structure for $\eta=0.7$.
We employ the standard notation for the high symmetry points in the fcc Brillouin zone.
The third and fourth bands are crossing between the X and W points.
(b)-(e) Weyl points for the isotropic case. The tension parameters are
(b) $\eta=0.74$, (c) $\eta=0.67$, (d) $\eta=0.66$ and (e) $\eta=0.58$.
The three blue arrows marked with $\bm k_1$, $\bm{k_2}$ and $\bm k_3$ are the reciprocal vectors of the diamond lattice. 
The dased lines indicate the line nodes in the chiral symmetric
 case, which are found irrespective of the value of $\eta$. (For the
 chiral symmetric case, there are other line nodes in a
 certain range of $\eta$. See Fig.~\ref{fig:WeylDet} and Ref.~\onlinecite{takahashi2017}.)}
\end{figure}

\section{\label{sec:pos_NNN}position of Weyl points}
\subsection{Isotropic case}
We first see a typical band structure in Fig.~\ref{fig:WeylIso}(a). For $\eta=0.7$, we see a Weyl point as a band crossing on the X-W line. 
Here, the crossing is between a flat band and a dispersive band, implying the found Weyl point is marginal in terms of the type I and type II classification.
Then, equipped with the method in the previous section, we
investigate Weyl points in NNN model of mechanical diamond more in detail.
Figures \ref{fig:WeylIso}(b)-\ref{fig:WeylIso}(e) shows Weyl points for an isotropic case, where $\kappa=1/(1-\frac{2}{3}\eta)$,
$\kappa'_i=0.2/(1-\frac{2}{3}\eta)$ ($i=1,2,...,6$), $\eta_0=1$, $\eta'_0=0.8$ and $m=2$. 
We have applied the dividing procedure twice.
In the first step, we use $N_B^{\text{1st}}=41$ and $N_D^{\text{1st}}=5$, and
in the second step, we use $N_B^{\text{2nd}}=5$ and
$N_D^{\text{2nd}}=3$, in which the length of the edge of the smallest
cube is 1/205 of the Brillouin zone size.
 A red (blue) point stands for a Weyl point of a monopole
(antimonopole) as described in Fig.~\ref{fig:WeylCalc}(b). 
We find six pairs of the Weyl points
moving from the W points to the X points
as the tension increases,
or $\eta$ decreases. Note that the line node in NN
model passes through the line connecting the W and X points.
In decreasing $\eta$, the Weyl points are created as
pairs at the W points, and annihilated with those from the other pairs at
the X points. 

\begin{figure}[tbp]
 \centering 
  \includegraphics[width=8.5cm]{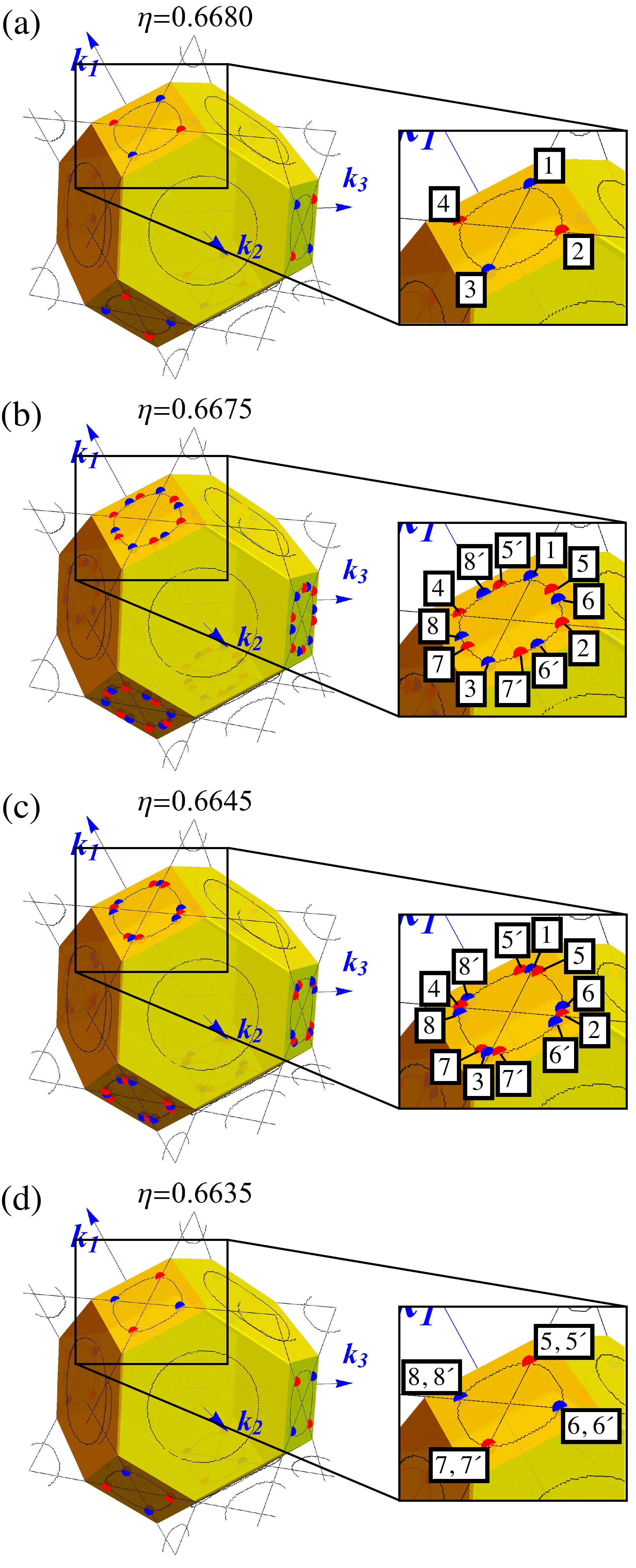}
\caption{\label{fig:WeylDet}Details for the switch of
 the positive and negative chiralities. The tension parameters are
(a) $\eta=0.6680$, (b) $\eta=0.6675$, (c) $\eta=0.6645$ and (d)
 $\eta=0.6635$. Black lines represent the numerically obtained line nodes in the NN model (i.e., with chiral symmetry) for each $\eta$.
}
%Some lines without intersecting do not have relations with the creation of the Weyl points.}
\end{figure}

Remarkably, the colors of the Weyl points are different
between Fig.~\ref{fig:WeylIso}(b) and Fig.~\ref{fig:WeylIso}(c), i.e.,
there is a rapid transmutation of the monopole charge from positive to
negative (or vice versa).
This transmutation occurs 
in between $\eta=0.67$ and $\eta=0.66$ \cite{footnote2}. Figure
\ref{fig:WeylDet} illustrates the detailed process of
the transmutation. 
Because of the three-fold rotational symmetry and the
reflection symmetry of the system, it is sufficient to examine one of
the square faces of the fcc Brillouin zone.
%In figure (b), twelve pairs of new Weyl points are created for $\eta=0.6675$.
In the rest of this paragraph, we focus on the number of Weyl points only on the one
square face. (The total number of generated Weyl points in the entire Brillouin zone is obtained by triplication.)
For an illustrative purpose,
we put numbers on found Weyl points as in the right panels of
Fig.~\ref{fig:WeylDet}. Also as a guidance, we draw line nodes for the
corresponding NN model with chiral symmetry in
Fig.~\ref{fig:WeylDet}. (There are no line nodes anymore without the
chiral symmetry.)
As $\eta$ decreases from 0.6680 to 0.6635, the number of
the Weyl points on the square face changes as
4$\rightarrow$12(original 4 + new 8)$\rightarrow$4. 
%The creation and annihilation of Weyl points occur in the same way at each corner
For $\eta=0.6680$, the four Weyl points
(No.~$1$--No.~$4$) are at the crossing points between the straight and
circularlike line nodes. 
For $\eta=0.6675$, new eight Weyl points
(No.~$5$--No.~$8$ and No.~$5'$--No.~$8'$) are found on the circularlike
line node. These new Weyl points are created as pairs of
No.~$5$-No.~$6$, No.~$6'$-No.~$7'$, No.~$7$-No.~$8$, and No.~$8'$-No.~$5'$.
For $\eta= 0.6645$, the two pairs of the monopoles No.~$5$-No.~$5'$ and No.~$7$-No.~$7'$ approach
to the antimonopoles No.~$1$ and No.~$3$ along the trace of the
circularlike line nodes, respectively. 
At the same
time, the two pairs of the antimonopoles No.~$6$-No.~$6'$ and
No.~$8$-No.~$8'$ come close to
the monopoles No.~$2$ and No.~$4$.
Finally, only two monopoles and two antimonopoles that
have the opposite charges from the $\eta=0.6680$ case
remain at the intersection of the
straight and circularlike line nodes for $\eta=0.6635$.

\begin{figure}[tbp]
 \centering 
  \includegraphics[width=8.5cm]{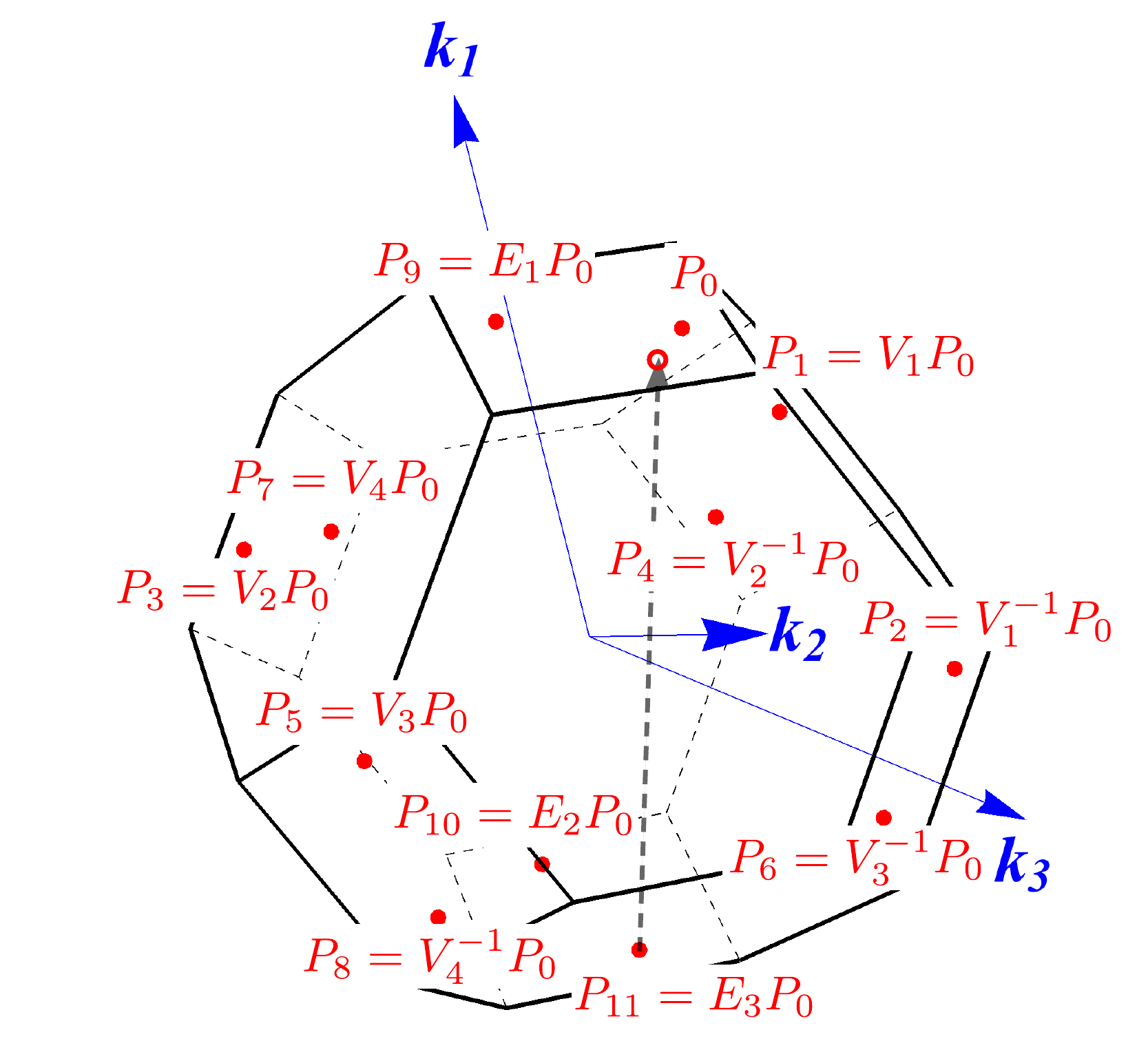}
\caption{\label{fig:GroTheory}Symmetric operations of
the tetrahedral symmetry in the fcc Brillouin zone.
The 12 points $P_i\,(i=0,1,...,11)$ correspond to the
generated Weyl points. A dashed vector is a guide to show that 
a pair of the Weyl points created on the zone boundary are actually identical.}
\end{figure}

It is helpful to think of the tetrahedral symmetry of
the NNN springs in understanding the number of the Weyl points in the
first Brillouin zone. Here, we use the numbers
in Fig.~\ref{fig:NNN}(b). 
There are $4!=24$ symmetric opetations to exchange the vertices in the tetrahedron, which are mapped on the permutations of (1,2,3,4) to form the symmetric group $S_4$. Out of 24, 12 do not involves the mirror operation (see Fig.~\ref{fig:GroTheory}), forming the alternating subgroup $A_4$, whose elements are: $I(1, 2, 3, 4)=(1, 2, 3, 4)$, $V_1(1, 2, 3, 4)=(2, 3, 1, 4)$,
$V_1^{-1}(1, 2, 3, 4)=(3, 1, 2, 4)$, $V_2(1, 2, 3, 4)=(1, 3, 4, 2)$,
$V_2^{-1}(1, 2, 3, 4)=(1, 4, 2, 3)$, $V_3(1, 2, 3, 4)=(3, 2, 4, 1)$,
$V_3^{-1}(1, 2, 3, 4)=(4, 2, 1, 3)$, $V_4(1, 2, 3, 4)=(2, 4, 3, 1)$,
$V_4^{-1}(1, 2, 3, 4)=(4, 1, 3, 2)$, $E_1(1, 2, 3, 4)=(2, 1, 4, 3)$,
$E_2(1, 2, 3, 4)=(3, 4, 1, 2)$ and $E_3(1, 2, 3, 4)=(4, 3, 2, 1)$. Then, if we have a Weyl point with a specific chirality at a \textit{generic point} in the Brillouin zone, $S_4$ generates 24 Weyl points in total. Among the 24, 12 generated by $A_4$ have the same chirality as the original one, while the other 12 generated by the operations involving the mirror operation have the opposite chirality as the original one. In contrast, a Weyl point on the \textit{high symmetric X-W lines} relocate to 12 (instead of 24) points by the permutation.

\begin{figure}[tbp]
 \centering 
  \includegraphics[width=8.5cm]{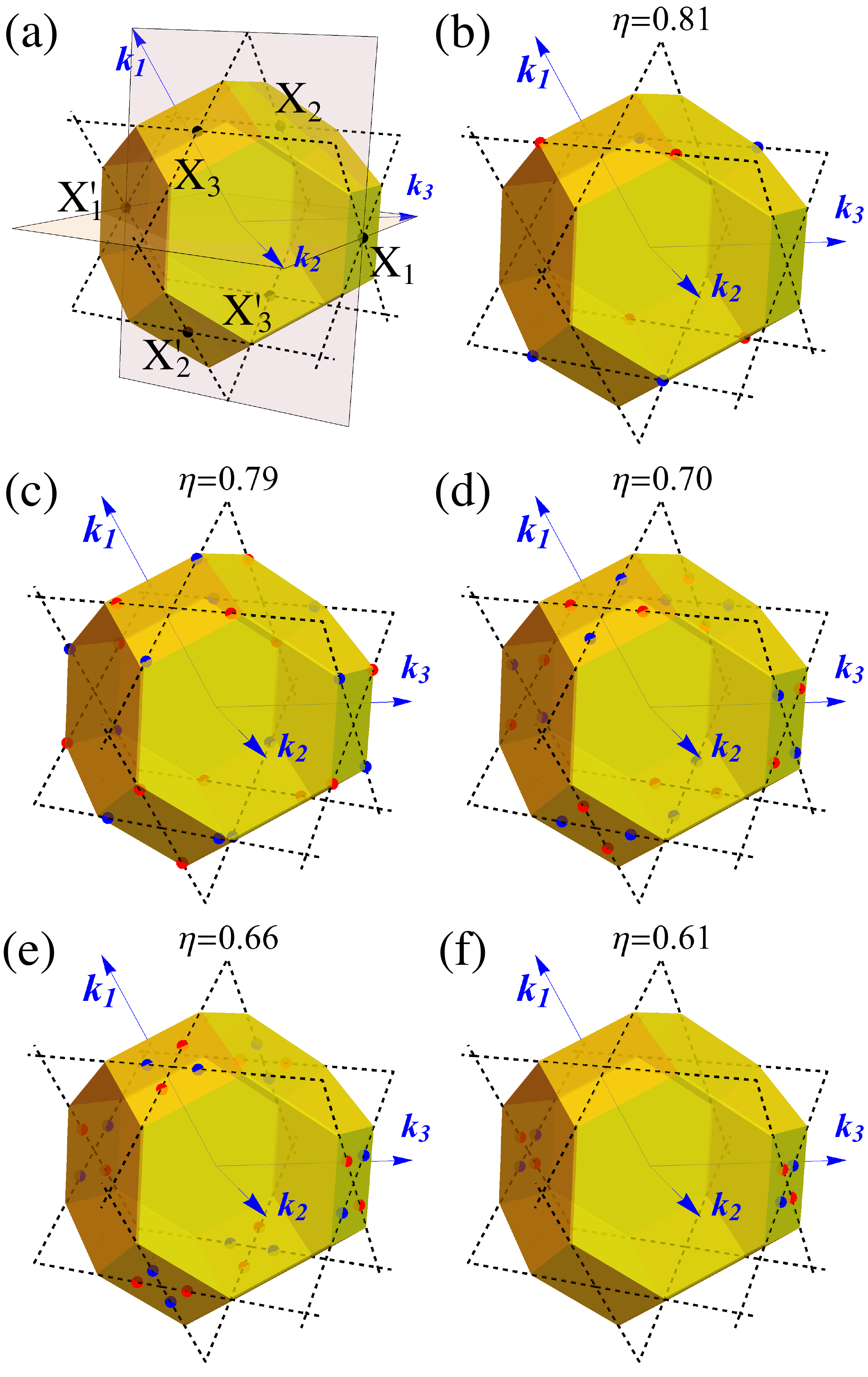}
\caption{\label{fig:AniWeyl}(a) Notations for the high
 symmetric points in the Brillouin zone. 
$\mathrm{X}_i$ and $\mathrm{X'}_i \ (i=1,2,3)$ points are equivalent respectively.
The two planes indicate mirror planes in the Brillouin zone
remaining in the anisotropic case.
(b)-(f) Weyl points for the anisotropic case. The tension parameters are
(b) $\eta=0.81$, (c) $\eta=0.79$, (d) $\eta=0.70$, (e) $\eta=0.66$ and (f) $\eta=0.61$.}
\end{figure}

\subsection{Anisotropic case}
In the previous paragraph, we have seen that the number
of the Weyl points are restricted by the tetrahedral symmetry.
Here, let us introduce anisotropy to break this symmetry.
In the isotropic case
discussed so far, we have been using $\eta'/\eta=0.8$ for all the NNN
springs. In order to induce anisotropy, we modify $\eta'/\eta$ for the
springs along $\bm{R}^{(1)}_1$ direction to 1. The system does not have
tetrahedral symmetry anymore, but again, it should be noted that the
forces on a single mass point cancel out in a pairwise manner in
equilibrium, and therefore, the applied modification is not harmful in
keeping the ideal diamond structure.
Figure \ref{fig:AniWeyl} shows Weyl points for the anisotropic case.
Due to the symmetry breaking,
timing of the creation and the annihilation
differs from the isotropic case.
As we decrease $\eta$, two pairs of
Weyl points are generated in $k_1=\pm \pi$ plane
at $\eta=0.81$. Then, other four pairs emerge
at $\eta=0.79$. (Remember that in the isotropic case,
the six pairs are created simultaneously.)
All six pairs switch the chirality
in between $\eta=0.70$ and $\eta=0.66$, although
the switch does not take place simultaneously. The switch of
the four pairs near $\mathrm{X}_2$, $\mathrm{X'}_2$,
$\mathrm{X}_3$ and $\mathrm{X'}_3$ points is first
and that of the two pairs near $\mathrm{X}_1$ and $\mathrm{X'}_1$
points is second.
Finally, only two pairs near the
$\mathrm{X}_1$ and $\mathrm{X'}_1$ points remain for $\eta=0.61$.

As we have seen, in the isotropic case, if a Weyl point
is generated on X-W lines, the symmetry automatically gives 12 Weyl
points in total. In the anisotropic case, the multiplicity of the Weyl
points obeys a different rule.
The anisotropy reduces the symmetry from $A_4$ to $C_2$, and even if it is combined with the mirror operation, a Weyl point at a generic point is copied onto only 4
points. (Note that 4 is the expected minimum number of Weyl points with the time-reversal symmetry.) Indeed, we find 4 Weyl points for $\eta=0.81$, and another 8 Weyl points are added afterward as $\eta$ decreases to $0.79$, and so on. 
The key is that the symmetry breaking reduces the number of simultaneously created pairs, in consistent with the group theory.

\section{\label{sec:Fer} Bulk section Chern number and Fermi arc}
\begin{figure}[tbp]
 \centering 
  \includegraphics[width=8.5cm]{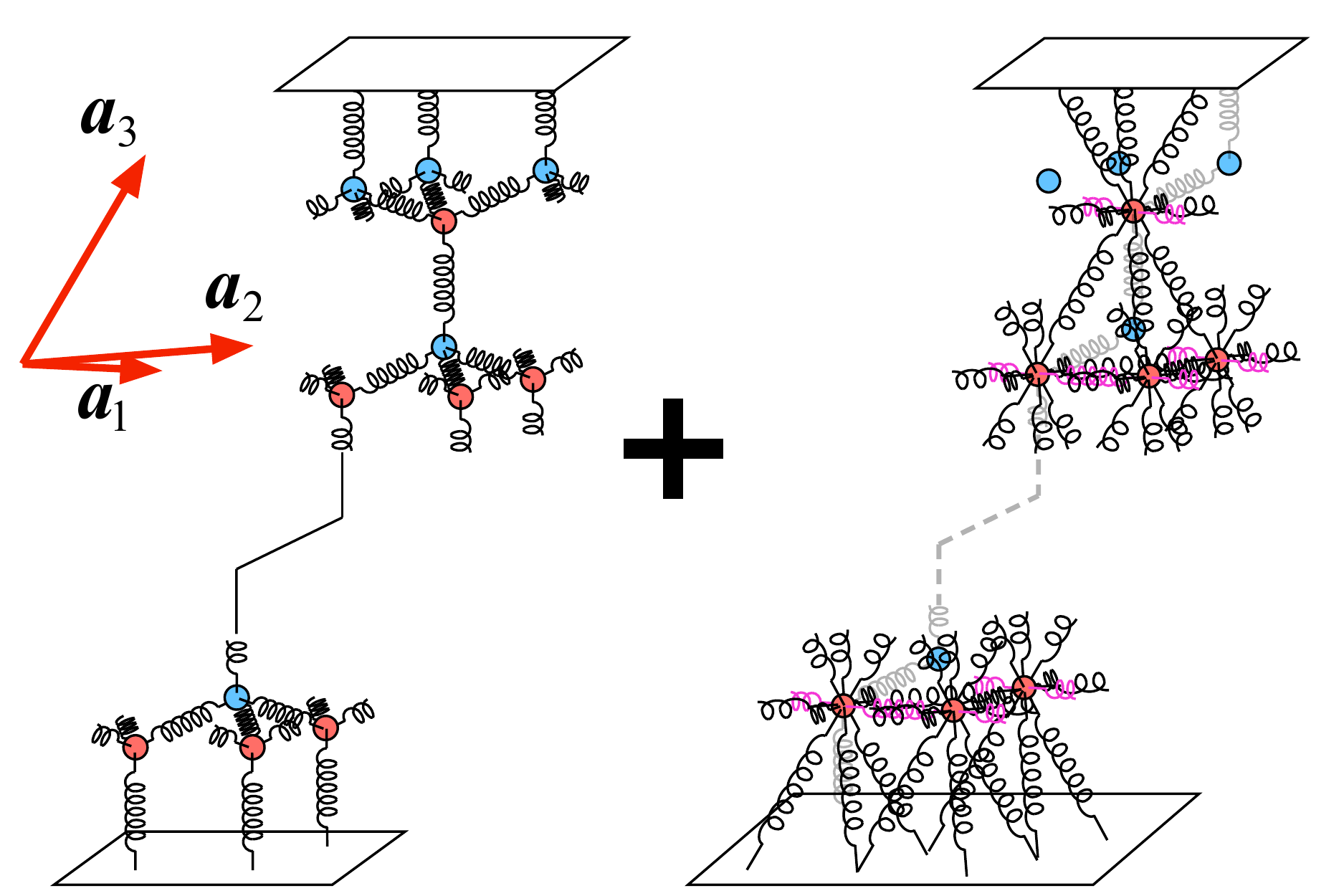}
\caption{\label{fig:BouCon}Schematic picture of the NNN
 system with a boundary. The tension parameter for the magenta springs,
 which are arranged to be parallel to the surface, is different from the one
 for five other NNN springs.} 
\end{figure}

 In this section, we show chiral edge states (or Fermi
 arc in this case) characterized by a bulk topological number. Let us
 start with the definition of the bulk topological number. In our 3D
 model, we have section Chern number (SCN)
\cite{Avron1983,Halperin1987,Kohmoto1992,Hatsugai2004}, whose definition is
\begin{equation}
 C(k_i)=-\frac{\mathrm{i}}{2\pi}\int_S \mathrm{Tr} \, \mathrm{d}\mathcal{A}\Bigr|_{k_i:\mathrm{fixed}}.
\end{equation}
Here, $\mathcal{A}$ is a non-Abelian Berry connection
$\mathcal{A}=\psi^{\dagger}\mathrm{d}\psi$.
The state $\psi$ is a triplet formed by
three eigenstates with lower frequencies. 
The integration is taken over the two-dimensional Brillouin zone with
$k_i$ ($i=1,2,3$) fixed. 
In the isotropic case, the SCN defined in this way is always trivial for any fixed momentum
because of mirror planes
passing through the $k_1$-, $k_2$-, and $k_3$-axes.
In the anisotropic model, on the other hand, $C(k_2)$
and $C(k_3)$ can be nontrivial in a certain range of fixed momentum due
to the absence of a mirror plane through the $k_2$- and $k_3$-axes. 
This nontrivial SCN is related to the surface states on the specific surface in Fig.~\ref{fig:BouCon}. Actually, the SCN can be defined on different types of 2D slices of the 3D Brillouin zone. For instance, it is in principle possible to prepare the SCN isolating the Weyl points in the isotropic case.

\begin{figure*}[tbp]
 \centering 
  \includegraphics[width=17cm]{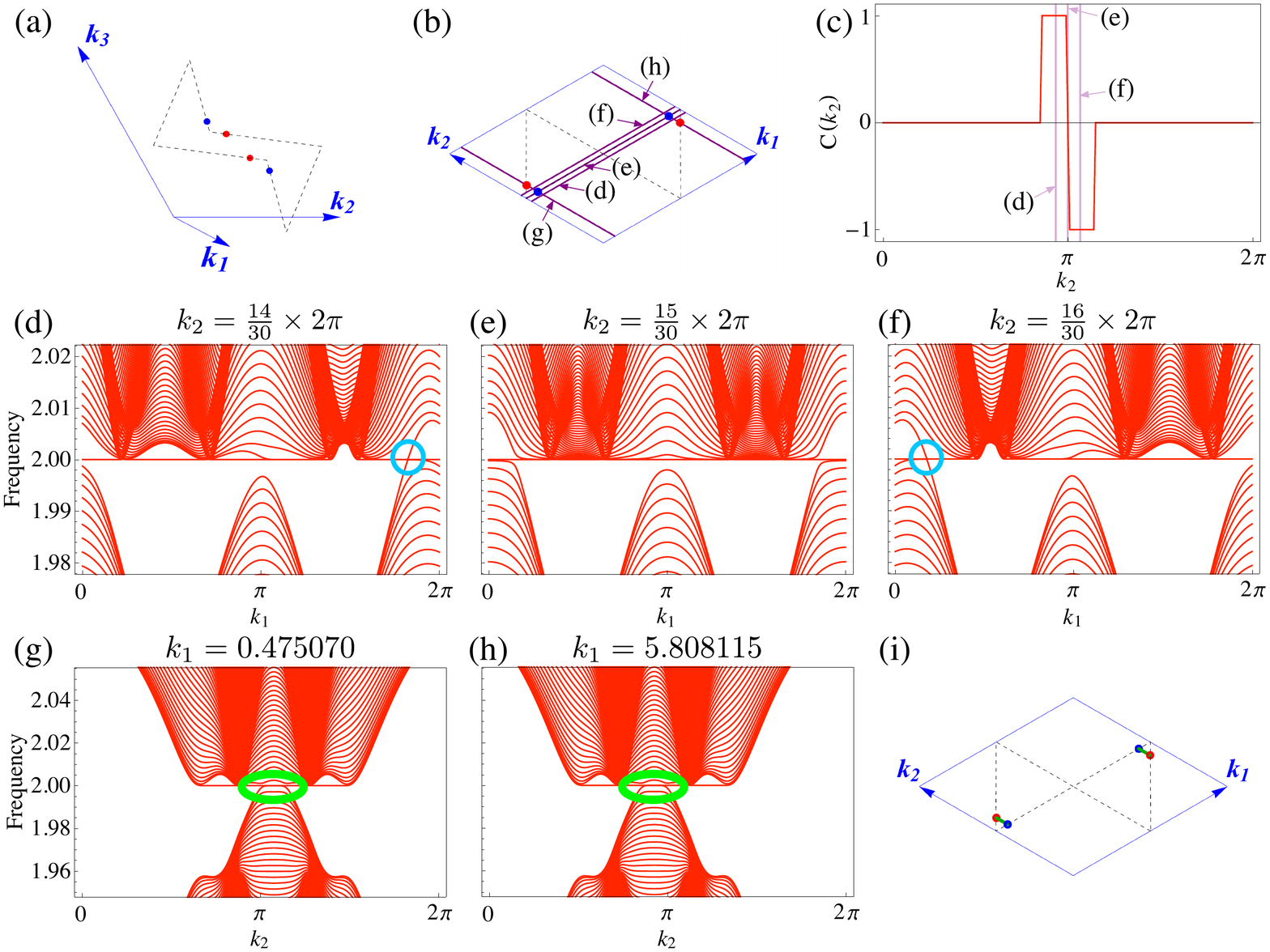}
\caption{\label{fig:FerArc}(a) Weyl points of the
 anisotropic model for $\eta=0.61$ in the parallelepiped Brillouin
 zone. (b) Weyl points projected onto the $k_1$-$k_2$ plane.
 (c) SCN of this model in $k_2$ direction.
 (d)-(h) Surface band structures along the five purple lines named
 (d)-(h) in (b), respectively. In (d)-(f), $k_2$ is fixed to 
 (d) $k_2=\frac{14}{30}\times 2\pi$, (e) $k_2=\frac{15}{30}\times 2\pi$
 and (f) $k_2=\frac{16}{30}\times 2\pi$, and the band structures are
 scanned by $k_1$. In the region highlighted by an aqua circle, we see dispersive and flat bands. The dispersive one connecting the lower and the upper bands is the chiral edge mode. The flat one has also the surface origin.
 In (g) and (h), $k_1$ is fixed to (g) $k_1=0.475070$
 and (h) $k_1=5.808115$, and the band structures are scanned by $k_2$.
 (The values of $k_1$ are numerically determined to hit the projected
 Weyl points.) A Fermi arc is marked by the yellow-green ovals.
 (i) A Fermi arc projected onto the $k_1$-$k_2$ plane, drawn as green lines.}
\end{figure*}

To obtain surface states, we consider a system that is periodic in $\bm a_1$ and
$\bm a_2$ direction but finite in $\bm a_3$ direction. In the $\bm{a}_3$ direction,
the system is terminated with the fixed boundary condition as illustrated in
Fig.~\ref{fig:BouCon}. (When we are hanging the entire system in a cage, this corresponds to set one face of the cage along $\bm{a}_1$-$\bm{a}_2$ plane, and focus on the motion of the mass points near that face.)
The surface states with this boundary condition for the NN model has been discussed in our
previous paper \cite{takahashi2017}.
Now for the NNN model, Fig.~\ref{fig:FerArc}
shows $C(k_2)$ and surface states of a system with $400$ layers
in $z$ direction for $\eta=0.61$.
As expected, we see chiral edge states connecting
the third and the fourth lowest bands for $k_2$ with finite $C(k_2)$.
$C(k_2)$ changes when
the plane of fixed $k_2$ crosses a gapless point
with topological charge, i.e., Weyl point. 
One monopole (antimonopole) adds $+1$ ($-1$) to the SCN. We also notice
that the number of the chiral edge states matches to the SCN, with the sign of the SCN related to the sign of the slope of the chiral edge states.
Note that flat bands that is already found in the NN model~\cite{takahashi2017}
remain even in the NNN model. 
The Fermi arc is shown in Fig.~$\ref{fig:FerArc}$(i), which
has no dependence on wave number in the
$\bm k_1$ direction here.

\section{\label{sec:con}Concluding Remarks}
To summarize, we have investigated Weyl points of the mechanical diamond.
By introducing the NNN springs and the variation in the mass parameter, 
several pairs of the Weyl points in the frequency dispersion appear
in the Brillouin zone. 
The Weyl point positions are numerically obtained by
calculating topological charges. As a function of the
tension parameter, the configuration of the Weyl points in the Brillouin
zone shows an interesting evolution.
For the isotropic case with tetrahedral symmetry,
six pairs of the Weyl points move on the high-symmetric
W-X lines as the tension increases.
In a narrow range of the tension parameter, several
extra pairs of Weyl points are created and annihilated, and after the
annihilation, the chirality of the remaining Weyl points is flipped.
For the anisotropic case without tetrahedral symmetry,
the sequence of the creation/annihilation of Weyl points is a little
different from the one in the isotropic case. We also demonstrate the relation between the surface states and the bulk SCN in the case with anisotropy.

%Elaborating the design, for instance, to have Weyl points better isolated from the other part of the band is an interesting future task.
In a practical view point, there is still a room to improve our design. For instance, the frequency-momentum range where the Fermi arcs are clearly visible is narrow in our design. It is an interesting future task to propose an elaborated design to have Weyl points better isolated from the other part of the band. Alternatively, one may elaborate the experimental observation scheme to access the narrow range in the frequency-momentum space. In a two-dimensional mechanical system with nontrivial topology, the topological edge modes are excited, for instance, by poking a mass point by air emitted by a nozzle \cite{Nash24112015}. In principle, it is possible to improve the selectivity of the excited mode by using multiple nozzles and poking several mass points with appropriate phase shifts.

Finally, it is worth noting that the current 3D printing technique might be useful for fabricating 3D structures. In a recent theoretical paper \cite{Wang2018njp}, the usage of the 3D printer to realize Weyl points in a vibration spectrum in an elastic media is discussed. Notably, in Ref.~\onlinecite{Wang2018njp}, the Weyl points are first discussed in a tight-binding model, i.e., a discrete model, which can be exactly mapped to a spring-mass model, and then propose a design of a continuum elastic system, which is 3D printable and has a vibration spectrum similar to the one in a discrete model. Although the design in our proposal and the one in Ref.~\onlinecite{Wang2018njp} are very different, it opens an interesting possibility to print a system capturing essential features of our model. One might think that the parameter $\eta$ in our analysis is specific to the spring-mass model. However, as is pointed out in Ref.~\onlinecite{Kariyado2015}, $\eta$ actually controls the ratio between the longitudinal and the transverse coupling constants, and it is likely that a clever structural design can account for that.

\begin{acknowledgments}
 We thank C. Mudry for useful discussions.
 This work was partly supported by JSPS KAKENHI Grant Numbers JP17K14358
 (TK), JP16K13845 (YH), JP17H06138 and JP25107005 (YT, YH). 
\end{acknowledgments}


\begin{thebibliography}{38}
\bibitem [{foo({\natexlab{a}})}]{footnote1}%
  \BibitemOpen
  \href@noop {} {}\bibinfo {howpublished} {In our design, the forces exerted by
  the springs are balanced between themselves, but if we further consider the
  gravity, we may have to add extra springs to compensate it.}%
\bibitem [{foo({\natexlab{b}})}]{footnote2}%
  \BibitemOpen
\bibfield  {journal} {  }\href@noop {} {}\bibinfo {howpublished} {The
  transition occurs very close to 2/3, but we can shift this value by changing
  parameters in our model without breaking symmetry, and 2/3 is unlikely to be
  a special value.}%
\end{thebibliography}
\end{document}